\documentclass[smallextended]{svjour3}
\usepackage{amsfonts}
\usepackage{bm}
\usepackage{verbatim}
\usepackage{color}
\usepackage{graphicx}

\begin{document}

\title{$^3$He Universe 2020}

\author{G.E.~Volovik}

\institute{G.E.~Volovik \at
Department of Applied Physics, Aalto University, P.O. Box 15100, FI-00076
Aalto, Finland \\
Landau Institute for Theoretical Physics, acad. Semyonov av., 1a,
142432, Chernogolovka, Russia}

\date{Received: \today / Accepted: date}

\maketitle

\begin{abstract}
The latest news from $^3$He Universe are presented together with the extended map of the Universe.
\end{abstract}

\tableofcontents

\section{Introduction}

The old information on the $^3$He Universe can be found in Refs. \cite{VollhardtWolfle1990,Salomaa1987,Volovik2003},
The more recent information is in Refs.\cite{Mizushima2015,Mizushima2016,LeeHalperin,TopSuperReview}.

Here are the latest news from $^3$He Universe 2020.

%%%%%%%%%%%%%%%%%%%%%%%%%%%%%%%%%%%%%%%%%%%%%%%%%%%%%%%%%
%%%%%%%%%%%%%%%%%%%%%%%%%%%%%%%%%%%%%%%%%%%%%%%%%%%%%%%%%
\begin{figure}
\centerline{\includegraphics[width=1.2\linewidth]{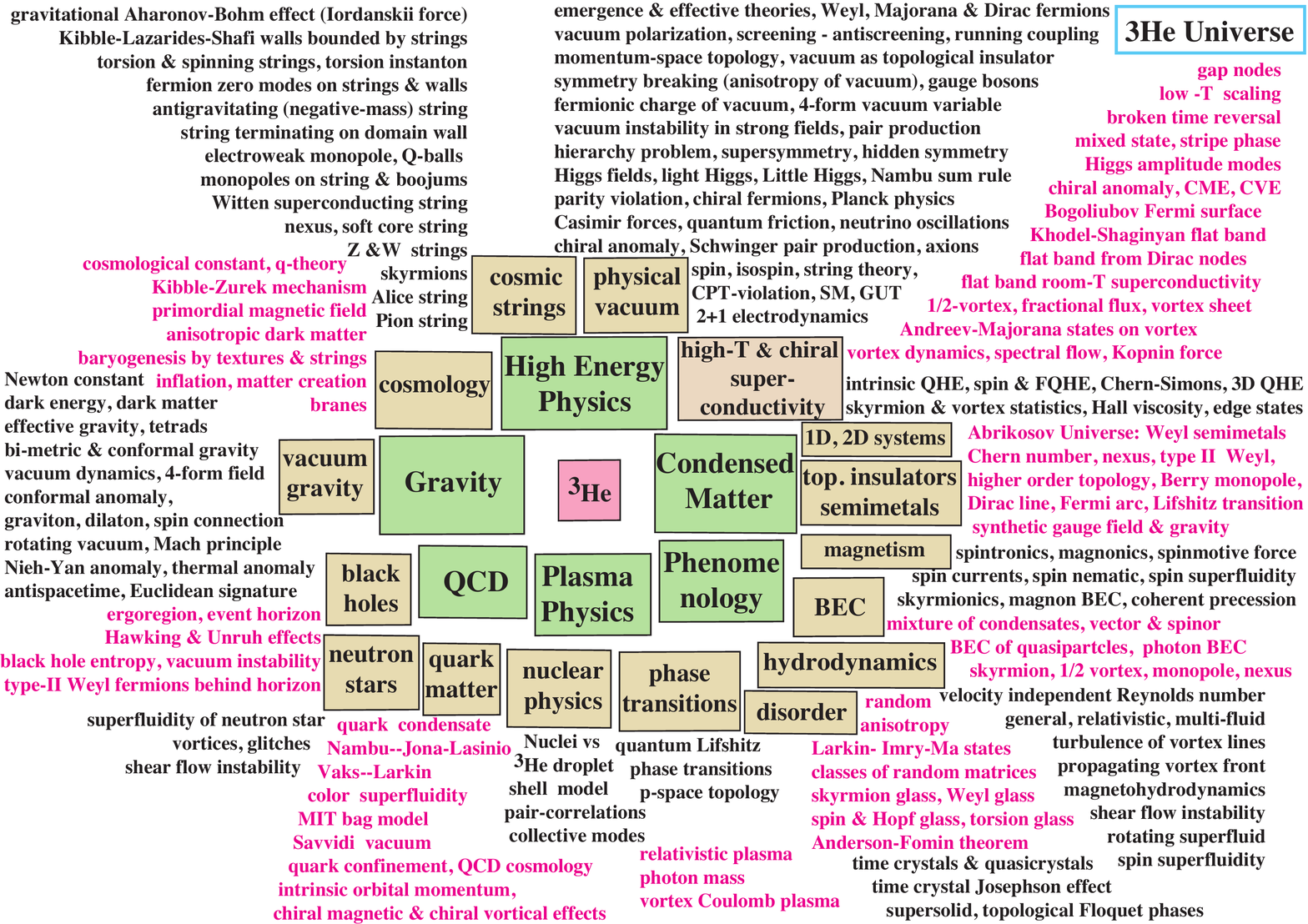}}
\caption{$^3$He Universe-2020. On the Far East is a connection with the neighboring Abrikosov Universe, which was also born in 1971 \cite{Abrikosov1971,Abrikosov1972}. The $^3$He Universe-1997 is in Fig. 1 of Ref.  \cite{Volovik1998}, and the comparison demonstrates inflating expansion of the $^3$He Universe.
}
\label{fig:2020}
\end{figure}
%%%%%%%%%%%%%%%%%%%%%%%%%%%%%%%%%%%%%%%%%%%%%%%%%%%%%%%%%
%%%%%%%%%%%%%%%%%%%%%%%%%%%%%%%%%%%%%%%%%%%%%%%%%%%%%%%%%

\section{Topology}

Superfluid phases of $^3$He opened the new area of the application of 
topological methods to condensed matter systems, see recent reviews in Refs.\cite{Mizushima2016,TopSuperReview}.

\subsection{Topological superfluids}

The phases of superfluid $^3$He are the best representatives of different families of topological materials. In bulk liquid $^3$He there are two topologically different superfluid phases \cite{VollhardtWolfle1990}.  One is  the chiral superfluid $^3$He-A with topologically protected Weyl points in the quasiparticle spectrum. In the vicinity  of the Weyl points, quasiparticles behave as Weyl fermions moving in the effective gauge and gravitational fields. Another phase is the fully gapped  time reversal invariant superfluid $^3$He-B. It has topologically protected gapless Majorana fermions living on the surface.  In $^3$He confined in the nematically ordered aerogel, the polar phase of $^3$He has been stabilized \cite{Dmitriev2015,HalperinParpiaSauls2018,Halperin2019}. It is the time reversal invariant superfluid, which contains Dirac nodal ring in the fermionic spectrum and flat band on the surface. 

\subsubsection{Chiral $^3$He-A, Weyl fermions, flat band of Majorana fermions on vortices}

Chirality of $^3$He-A has been probed  in the torsional oscillator measurements that distinguished between states of opposite chiralities
\cite{Krusius2012,Golov2012}. The topological manifestation of chirality is the separation of left-handed and right-handed Weyl points in momentum space. Due to bulk-vortex correspondence, the separation of the Weyl points leads to the flat band of Majorana fermions living in the vortex core \cite{KopninSalomaa1991,VolovikCore2011}.
 In topological Weyl semimetals the similar bulk-surface correspondence produces the so-called Fermi arc on the surface of the material \cite{Vishwanath2011}.

\subsubsection{$^3$He-B, higher order topology}

The topological superfluid $^3$He-B is the prototype of topological insulators and provides example of the higher order topology. The boundary of the B-phase contains 2D gapless 
Majorana fermions, which are supported by topology and symmetry of the B-phase. When magnetic field is applied, the time reversal symmetry of the B-phase is violated. The gapless fermions are not protected any more by topology: they become massive 2D Dirac fermions. However, the topology is not destroyed completely: the boundary states acquire their own  topology, where the topological charge is determined by the sign of the Dirac mass.
The line on the boundary, which separates the surface domains with opposite signs of mass term, contains its own topologically protected 1D gapless Majorana fermions \cite{Volovik2010b}. The presence of such lines on the boundary of $^3$He-B is seen in NMR experiments \cite{Saunders2010}.

Such composite  3D-2D-1D correspondence characterizes the so-called second order topology \cite{Benalcazar2017}. In topological insulators and superconductors the third order topology is also possible, with 3D-2D-1D-0D  correspondence \cite{3rdOrder}. This may lead to the zero dimensionless Majorana modes in the corners of superconductor.

\subsubsection{Polar phase: Dirac nodal line and flat band}

Polar phase is the nodal line superfluid, which is similar to cuprate $d$-wave superconductors and nodal line semimetals. 
In both systems the Dirac nodal line in the spectrum is supported by topology: the $\pi$ change of the Berry phase along the loop around the nodal line. There are several important consequences of the Dirac line. 

One of them is the existence of the flat band (or approximate flat band -- drumhead states) on the surface due to the bulk-surface correspondence. The phenomenon of flat band 
is important for search of room-$T$ superconductivity because of the singular density of states \cite{Khodel1990,Kopnin2011,Heikkila2011}. The transition temperature $T_c$ is not exponentially suppressed as in conventional metals, but is the linear function of the coupling in the Cooper channel.  $T_c$ is proportional to the volume of the flat band, if the flat band is formed in the bulk \cite{Khodel1990}, or to the area of the flat band if it is formed on the surface of the sample \cite{Kopnin2011,Heikkila2011}. In nodal line semimetals the area of the flat band is determined by the projection of the nodal line to the surface of the sample. The largest area is obtained when the nodal lines move to the boundaries of the Brillouin zone, where they cancel each other, and the nodal line semimetal is transformed to the topological insulator with the surface flat band \cite{Heikkila2020}.

\subsubsection{Bogoliubov Fermi surface}

The other important consequence of the nodal line takes place in superfluids and superconductors in the presence of supercurrent.
Supercurrent  violates parity and time reversal symmetries, as a result the Dirac line in the spectrum of Bogoliubov quasiparticles transforms to the Fermi surface of quasiparticles within the superconducting state -- the so-called Bogoliubov Fermi surface
\cite{Volovik1989e,Agterberg2018,Hirschfeld2020}. 

In the moving polar phase of $^3$He, the Bogoliubov Fermi surface has an exotic shape: it consists of two Fermi pockets which touch each other at two pseudo-Weyl points \cite{Autti2020}. In cuprate superconductors, the local Bogoliubov Fermi surfaces caused by supercurrents around  Abrikosov  vortices lead to the $\sqrt{H}$ field dependence of the electronic density of states in the vortex state of superconductor \cite{Moler1997}.

\subsection{Quantum anomalies}

The nontrivial topology of the superfluid phases of $^3$He lead to different types of quantum anomalies; chiral anomaly, gravitational anomaly, and mixed anomalies.

\subsubsection{Chiral anomaly, angular momentum paradox}

Chiral anomaly is the anomalous production of fermions from the vacuum, which is described by the Adler-Bell-Jackiw equation \cite{Adler1969,BellJackiw1969,Adler2005}. The anomalous non-conservation of the chiral current has been  verified in experiments with vortex-skyrmions in rotating $^3$He-A  \cite{BevanNature1997}. The anomalous  production of  Weyl fermions by moving skyrmions  leads to the anomalous production of the linear momentum, and thus to the extra force, which acts on the vortex. This spectral-flow or Kopnin force is measured, which allows to extract the fundamental prefactor in the Adler-Bell-Jackiw equation, which depends only on symmetry and fermionic content of the quantum vacuum. The measured integer number is in the full agreement with the Adler-Bell-Jackiw equation applied to Weyl fermions in $^3$He-A  (two left-handed fermions at one Weyl point and two right-handed fermions at another Weyl point).

In Weyl semimetals the manifestation of the chiral anomaly in experiments is not so spectacular, it leads to negative magnetoresistance, when the magnetic field is parallel to the current \cite{Huang2015}. 

Chiral anomaly   solves the paradox of the orbital angular momentum in chiral superfluids. The deviation of the orbital momentum from its natural  value,
$L_z=\hbar \nu N/2$ (where $N$ is the number of particles, and $\nu$ is the angular momentum of Cooper pair)
is fully determined by the spectral flow either in bulk or on the surface of the superfluid, see recent papers \cite{Tada2015,Volovik2014}.
The similar spectral asymmetry in the vortex core leads to the modification of the angular momentum of quantized vortices \cite{Moroz2017}.

 The paradox of the orbital angular momentum in chiral superfluids may have something common with the proton spin puzzle, however the present understanding suggests that the chiral anomaly effects are too small to explain the ‘spin crisis’ \cite{Brodsky2019}.
 
 Another phenomenon related to the orbital angular momentum is the Hall viscosity -- the non-dissipative response of stress tensor to the velocity gradients. 
 In the Eq.(9.81) of the book \cite{VollhardtWolfle1990} six Hall viscosity terms in $^3$He-A with coefficients $\gamma_\perp$, $\gamma_\parallel$ and $\beta^{\rm R}_a$ with $a=1,2,3,4$ are presented. The Hall viscosity in the 2D chiral superfluids is now under discussion in the literature, see review \cite{Rose2020}. The quantization of the Hall viscosity in terms of the orbital momentum  is suggested. However, one may expect that the spectral flow on the boundaries of the 2D system leads to the same reduction of the effect as in the paradox of the orbital  momentum.
 
\subsubsection{Chiral magnetic and chiral vortical effects}
\label{CMVE}

The interplay of quantum anomalies with magnetic field and vorticity results in a variety of non-dissipative transport phenomena in systems with chiral fermions \cite{Kharzeev2016}. This is popular for consideration of different effects in the quark–gluon plasma
created in relativistic heavy ion collisions. The Chiral Magnetic Effect (CME) and the Chiral Vortical Effect (CVE) describe the generation
of non-dissipative electric current along an external magnetic field or along the vortex. They are described by the topological quantum numbers, similar to that, which operate in the intrinsic quantum and spin Hall effects.

The experimental signature of the CME in $^3$He-A is the helical instability of the superflow generated by the Chern–Simons term, expressed via effective gauge fields acting on Weyl fermions \cite{Krusius1998,Volovik2017}, $S\sim \int \mu_5 {\bf A}\cdot{\bf B}$. This CME is linear in the chiral chemical potential $\mu_5$. In $^3$He-A  the effective chiral chemical potential is formed by superflow, see Sec. \ref{NYsec}. The corresponding term in the free energy is linear in the superluid velocity, which leads to instability of flow towards creation of vortex-skyrmions (continuous doubly quantized vortices).

CVE is manifested by the current along the vortex, which is concentrated in the core of vortex-skyrmion \cite{Volovik2017b}.
It is important that the total current is zero: the current along a given vortex in the vortex lattice is compensated either by the countercurrent
in the core of another vortex, or by the countercurrent in the bulk. This supports the  Bloch theorem (see e.g. Ref. \cite{Yamamoto2015}), which prohibits the total current in the equilibrium state. 

Experimental observation of CVE in the A-phase is still waiting in the wings. The same is with the Chiral Separation Effect (CSE), which is dual to the CME with $S\sim \int \mu {\bf A}_5\cdot{\bf B}$, see e.g. \cite{Gorbar2015,SuleymanovZubkov2020}.

\subsubsection{Nieh-Yan gravitational anomaly}
\label{NYsec}

The relativistic nature of the Weyl fermions in the A-phase is explicitly manifested in the thermal contributions of Weyl fermions to the free energy at low temperature, $T\ll T_c$. In particular, three different $T^2$ terms in the gradient energy of $^3$He at low $T$ can be rewritten in the fully relativistic form:
\begin{equation}
F= \frac{T^2}{144}\sqrt{-g} {\cal{R}} -\frac{T^2}{6} \sqrt{-g} \mu_5^2 - \mu_5 \frac{T^2}{24}  \epsilon^{ijk} e_{i a}\mathcal{T}^a_{jk} \,.
\label{FT2}
\end{equation}
Here  $\mu_5$ is the chiral chemical potential of Weyl fermions;
  $e_{ia}$ are tetrads, which describe the effective spacetime in which Weyl fermions are moving,  ${\cal{R}}$ is the effective scalar curvature of this spacetime, and $\mathcal{T}^a_{jk}$ is the analog of torsion field \cite{Nissinen2020}.
  
   The first term in Eq.(\ref{FT2})  describes the temperature correction to the inverse Newton "constant" $1/G$ in the effective gravity.  The last term in Eq.(\ref{FT2}) is the manifestation of the Nieh-Yan gravitational anomaly, which is expressed  in terms of torsion field \cite{Nissinen2019}:
 \begin{equation}
\partial_{\mu} j^\mu_5 = -  \frac{T^2}{48}  \epsilon^{\mu\nu\alpha\beta} \mathcal{T}_{a\mu \nu}\mathcal{T}^a_{\alpha\beta}
 \,.
\label{jNiehYan}
\end{equation}
As distinct from the unknown ultraviolet parameter $\Lambda^2$ in the conventional torsional Nieh-Yan anomaly \cite{NiehYan1982a,NiehYan1982b,Nieh2007} (the ultraviolet parameter for the Nieh-Yan anomaly in $^3$He-A see 
in Refs. \cite{Nissinen2020b,Laurila2020}),
the thermal Nieh-Yan term contains  $T^2$ and thus is well defined. The prefactor in this term is funamental, being determined by the geometry, topology and number of  chiral quantum fields in the system. For the effective quantum relativistic fields in $^3$He-A  this parameter is $1/48$, while the other two parameters in Eq.(\ref{FT2}) are $1/6$ and $1/144$ \cite{VolovikZelnikov2003}.

In the $^3$He language,  the chiral chemical potential of Weyl fermions  is represented by the Doppler shift 
 $\mu_5=p_F{\hat{\bf l}}\cdot {\bf v}_{\rm s}$, where $p_F$ is Fermi momentum, $\hat{\bf l}$ is the unit vector along the angular momentum of Cooper pairs, $\pm p_F{\hat{\bf l}}$ are positions of two Weyl points and ${\bf v}_{\rm s}$ is superfluid velocity.
The Ricci scalar ${\cal{R}}$ and torsion $\mathcal{T}^a$ in the effective gravity are expressed in terms of the gradients of the order parameter (${\bf v}_{\rm s}$ and $\nabla\times {\hat{\bf l}}$)  \cite{Nissinen2020}.
 
\subsection{Composite topological objects}

Due to the multi-component order parameter which characterizes the broken $SO(3)\times SO(3)\times U(1)$ symmetry in superfluid phases of $^3$He, there are many inhomogeneous objects -- textures and defects in the order parameter field -- which are protected by topology and are characterized by topological quantum numbers. Among them there are quantized vortices, skyrmions and merons, solitons and vortex sheets, monopoles and boojums, etc. There are also composite topological objects, which combine defects  of different dimension. Among them there are Alice strings with soliton tail and analog of Kibble-Lazarides-Shafi cosmic walls terminated by Alice strings \cite{Kibble1982}, see recent review in Ref. \cite{Volovik2020c}. 

\subsubsection{Alice strings terminating solitons}

Half-quantum vortices (analogs of Alice string in cosmology) have been suggested more than 40 years ago, but have been observed in superfluid $^3$He only recently, first in the polar phase  \cite{Autti2016} and then in the A-phase \cite{Makinen2019}.
 Half-quantum vortex (HQV) itself represents the combination of the linear objects: it is partly a vortex 
 (the vortex with half of circulation quanta) and partly a spin vortex (the vortex with $\pi$ change of spin vector) \cite{Sauls2016}.
As a spin vortex, it is influenced by the spin-orbit interaction. As a results in the A-phase the HQV  is always accompanied by the solitonic tail, i.e. it becomes the termination line of the topological spin soliton which makes it energetically unfauvorable compared to other vortices (singly quantized vortex and vortex skyrmion). This was the reason, why it was so difficult to stabilize half-quantum vortices in the A-phase. In the polar phase the solitonic tails are absent if the magnetic field is along the nafen strands, and the half-quantum vortices become energetically favourable. They can be created either by cooling through $T_c$ under rotation or by fast cooling through $T_c$ without rotation, when the topological defects are formed by Kibble-Zurek mechanism \cite{Rysti2019}. 
 
 When  half-quantum vortices are created, they are pinned by nafen strands, and we can do with them whatever we want.
 If we tilt the magnetic field with respect to strands, the solitons appear, which are terminated by half-quantum vortices. Then by measuring the intensity of the satellite peak in the NMR spectrum, which comes from soliton, we can find the total length of solitons, and thus the total number of half-quantum vortices in the cell.
 
 We can make the phase transition from the polar phase to the $^3$He-A, and the half quantum vortices are still there. Moreover, we can make  the phase transition from the polar phase to the $^3$He-B, where half-quantum vortices cannot exist as topological objects. But again they remained  pinned \cite{Makinen2019}.

\subsubsection{Kibble-Lazarides-Shafi walls bounded by strings}

The object, which is formed in $^3$He-B, after the transition from the polar or A-phase with the pinned half-quantum vortices, is the domain wall terminated by pinned vortices \cite{Makinen2019,Zhang2020}. This composite object is the exact analog of the Kibble-Lazarides-Shafi wall bounded by cosmic strings in cosmology \cite{Kibble1982}. 

\subsubsection{Nexus}

The other composite objects can be constructed and pinned by nafen strands, including nexus (monopole or hedgehog, which connects two or more strings \cite{Zhang2020}), necklaces  \cite{Lazarides2019} and lattices of composite objects \cite{Volovik2020c}. But at the moment these analogs of Nambu monopoles \cite{Nambu1977} and their further extensions are still not resolved in NMR experiments.

Randomly pinned topological objects can provide different types of topological glasses.

 \subsection{From topological classes to topological glasses}
 
The quenched random
anisotropy provided by the confining material strands produces several different
glass states resolved in NMR experiments  in the chiral superfluid
$^3$He-A and in the time-reversal-invariant polar phase. The smooth
textures of spin and orbital order parameters in these glasses can be
characterized in terms of the randomly  distributed  topological charges,
which describe skyrmions, spin vortices and hopfions. In addition, in
these skyrmion glasses the momentum-space topological invariants are
randomly distributed in space. The Chern mosaic \cite{Ojanen2016}, Weyl glass, torsion
glass and other exotic topological sates are examples of close
connections between the real-space and momentum-space topologies in
superfluid $^3$He phases in aerogel, see review in Ref.\cite{Volovik2019}.

\subsubsection{Larkin-Imry-Ma state as Weyl glass}

One of the most spectacular discoveries made in superfluid $^3$He
confined in a nanostructured material like aerogel or nafen was the observation of the destruction of the long-range orientational order by a weak random  anisotropy \cite{Dmitriev2010,Askhadullin2014}, the so-called Larkin-Imry-Ma state \cite{Larkin1970,ImryMa,Volovik1996,Volovik2008}.
In the chiral A-phase it is the orbital vector ${\hat{\bf l}}$, which looses the long range orientational order. 
According to Mermin-Ho relation the disordered texture of the ${\hat{\bf l}}$-vector generates the random distribution of superfluid velocity. This state has random distribution of $\pi_2$ and $\pi_3$ topological charges, which describe skyrmions and hopfions correspondingly. Thus the Larkin-Imry-Ma states can be realized in the form of the skyrmion glass \cite{Chudnovsky2018} or/and hopfion glass.

  On the other hand,  the orbital vector ${\hat{\bf l}}$ determines the position of two Weyl points in momentum space. That is why the skyrmion glass also represents the topological Weyl glass, which is different from the conventional Weyl disorder with random shifts in the position of Weyl nodes \cite{Altland2018}.

\subsubsection{Anderson-Fomin theorem}

According to the Anderson theorem \cite{Anderson1959}, the $s$-wave superconductors with non-magnetic impurities are robust to weak disorder, i.e. the critical temperature $T_c$ is the same as in the clean limit. 

For a long time it was presumed that the Anderson theorem is not applicable to spin triplet superconductors, or to superconductors with nodes in the gap.  However, it was found by Fomin \cite{Fomin2018}, that if impurities in the polar phase have the form of infinitely long non-magnetic strands, which are straight and parallel to each other, the transition temperature also coincides with that in the clean limit. The reason is that  in the presence of columnar defects the polar phase can be considered as a set of independent 2D superfluids. The behavior of each 2D superfluid in the  presence of the corresponding 2D defects is similar to that of the $s$-wave superconductors.
This robustness of the polar phase to the columnar disorder is the main reason why the polar phase survives in nafen even for strong disorder, when all the other phases are removed from the phase diagram. 

Unusual properties of the polar phase in nafen have been discussed in Refs.  \cite{Autti2020,Eltsov2019b}.

 \section{Gravity and cosmology}

\subsection{Tetrad gravity, pregeometry and dimensionless physics}

In $^3$He Universe, gravity emerges together with the gauge fields in the vicinity of the topologically stable Weyl point. In $^3$He-A, gravity emerges in the form of tetrads, which are obtained as the spacetime dependent parameters of  expansion of the Green's function of Bogoliubov quasiparticles in the vicinity of Weyl points. 
This suggests that in our Universe, gravity and gauge fields are also the emergent phenomena, which come from the topology of the quantum vacuum.

 The $^3$He Universe suggests also the alternative scenario of the origin of gravity:  in $^3$He-B the tetrad field emerges as a composite field. Such origin of geometry of spacetime (the pregeometry), has been first discussed by Akama \cite{Akama1978} and more recently by Diakonov \cite{Diakonov2011,VladimirovDiakonov2012,VladimirovDiakonov2014}, where the tetrad field emerges as bilinear combination of the fermionic fields: 
\begin{equation}
e^a_\mu=i \left<\psi^\dagger \gamma^a \nabla_\mu \psi + \nabla_\mu\psi^\dagger \gamma^a  \psi \right>
\,,
\label{bilinear}
\end{equation}
where $\gamma^a$ are Dirac matrices.

This mechanism was discussed in Ref.\cite{Volovik1990} in terms of the symmetry breaking scenario, when  two separate Lorentz groups of coordinate and spin rotations are spontaneously broken to the combined  Lorentz symmetry group, 
$L_{L}\times L_{S}\rightarrow L_J$, and the tetrad field $e^a_\mu$ serves as the order parameter of the transition. 
This is the analog of the broken spin-orbit symmetry $SO(3)_{L}\times SO(3)_{S}\rightarrow SO(3)_J$ introduced by  Leggett for $^3$He-B \cite{Leggett1973}. The order parameter in  the B-phase, $A_{\alpha i}= \Delta_B R_{\alpha i}e^{i\Phi}$, contains the matrix of rotation $R_{\alpha i}$, which connects spin and orbital degrees of freedom of the liquid. The order parameter $e^a_\mu$ connects the spin and orbital  degrees of the quantum vacuum, thus realizing the the extension of the B-phase condensed matter vacuum to the $3+1$ vacuum of our Universe.

According to Eq.(\ref{bilinear}), the tetrad field $e_\mu^a$ transforms as a derivative and thus has the dimension of inverse length.
This gives the unexpected consequence for gravity and actually for any other fields living in such geometry.  All the physical quantities, which obey diffeomorphism invariance, such as the Newton constant, the scalar curvature, the cosmological constant, particle masses, fermionic and scalar bosonic fields, etc., are dimensionless (see details in Refs. \cite{Volovik2020d,Volovik2020e}).

\subsection{Type-II Weyl, black and white hole horizons}

In the moving $^3$He-A the Weyl cone is tilted. When the flow velocity exceeds the effective speed of light, the Weyl cone is overtilted and the Fermi surface is formed \cite{Zubkov2014b}. In the modern language this corresponds to the Lifshitz topological transition between the type-I Weyl system with isolated Weyl points to the type-II Weyl system, where Weyl points  connect the Fermi pockets \cite{Volovik2016,Zhang2017}. 

The type-II Weyl fermions also emerge behind the event horizon of black holes. That is why the black hole horizon serves as the boundary between type-I and type-II quantum vacua.  In Weyl semimetals and Weyl superfluids  the analog of event horizon is formed at the  interface, which separates the regions of type-I and type-II Weyl points.   This allows us on one hand to probe the Hawking radiation using Weyl superfluids and semimetals \cite{Volovik2016,Zhang2017}.  On the other hand,  one can study the interior of the black hole using the  experience of condensed matter, where the ultraviolet physics is known \cite{Zubkov2018,Zubkov2018a}.  The effective (acoustic) metric which describes the black hole analog in condensed matter, is known in general relativity as the Painleve-Gullstrand metric \cite{Painleve,Gullstrand}.

 Using the junction of type-I and type II Weyl semimetals one may probe the more exotic horizon of the white hole \cite{Wilczek2020}. On the other hand using the acoustic (Painleve-Gullstrand) metric in general relativity one can study the transformation of the black hole to white hole in the process of quantum tunneling \cite{Volovik2020BH}.

  \subsection{Routes to antispacetime and to Euclidean spacetime}
 
Using different types of the Weyl points one can model exotic spacetimes and the routes between them. For example
two roads to antispacetime exist in the presence of the the Kibble-Lazarides-Shafi wall bounded by strings in the B-phase \cite{Eltsov2019}: the safe route is around the Alice string (half-quantum vortex)). The dangerous route is across the domain wall. This dangerous route through the Alice looking glass is similar to the route of our Universe from spacetime to antispacetime via Big Bang.   In the A-phase the route to antispacetime takes place across the polar phase, where the metric of general relativity is degenerate \cite{NissinenVolovik2018}.

The transition from Minkowski spacetime to Euclidean space time, in which the signature of the metric changes, is also possible to probe in superfluids, using the collective modes of Bose-Einstein condensation (BEC) of magnons in the  polar phase \cite{Nissinen2017}.  Let us also mention that the magnon BEC -- the spontaneously formed coherent precession of spins -- provides the experimental realization  of time crystal and  time quasicrystal \cite{AuttiEltsovVolovik2018,AuttiHeikkinen2020}  following the Wilczek idea \cite{Wilczek2012}.

\subsection{Dark matter}

 Magnon BEC also provides the realization of the condensed matter analog of $Q$-balls \cite{AuttiHeikkinen2018}.
In relativistic quantum field theories, $Q$-ball is the nontopological soliton, which is stabilized by conservation of an additive quantum number. In our case it is the number of self-trapped magnons. In cosmology, $Q$-balls could have participated in baryogenesis,
formation of bosonic stars, and the dark matter.

\subsection{Dark energy, cosmological constant}

According to standard physics, the vacuum has an enormous energy density $\rho_{\rm vac}$. A positive contribution comes from the
zero-point energy of bosonic fields, such as electromagnetic field, and a negative contribution
comes from the fermionic fields -- from the so called Dirac vacuum, and there is no reason why they should cancel \cite{Weinberg1989}.
Again according to standard physics, $\rho_{\rm vac}$ should act as a gravitational source
-- effectively an enormous cosmological constant $\Lambda_{\rm vac}$. With the Planck scale providing a natural cutoff it is roughly 120 orders of magnitude
larger than is compatible with observations.
According to Bjorken, this is the oft-repeated mantra that "no one has any idea as to why the cosmological constant is so small" \cite{Bjorken2011}.

However, anyone who is familiar with superfluid $^3$He  at zero temperature and at zero pressure can immediately find a simple solution of the problem. The ground state of this superfluid is described by the thermodynamic potential $\epsilon(n) -\mu n$, where $\epsilon(n)$ is the energy density, $n$ is particle density and $\mu$ the chemical potential. According to the Gibbs-Duhem identity,  at zero temperature one has $\epsilon -\mu n=-P$, where $P$ is pressure.  That is why this  thermodynamic potential  is  equivalent to $\rho_{\rm vac}$, which obeys the equation of states characterizing the dark energy, $\rho_{\rm vac}=-P_{\rm vac}$. 

The low energy modes of superfluid $^3$He are described in terms of bosonic and  fermionic quantum fields (Higgs fields, Nambu-Goldstone fields and Bogoliubov fermions). One may think that zero-point energies of these modes contribute to $\epsilon(n)$. However this is not so, since the energy $\epsilon(n)$ fully comes from the ultraviolet physics, in our case from physics of the $^3$He atoms. The collective excitations of the liquid $^3$He contribute only to such effects, which depend on the low-energy (infrared) physics, such as the Casimir effect.   In the absence of environment, i.e. at zero pressure, the thermodynamic potential of the system is exactly zero,   $\epsilon(n) -\mu n=-P=0$. That is why  in the full equilibrium, the microscopic (Planck scale) energy $\epsilon(n)$ is cancelled by the counter-term $\mu n$ without any fine-tuning. This comes purely from thermodynamics, and does not depend on the phases of superfluid $^3$He (A-phase, B-phase, polar phase, etc.) and on the quantum fields living in these $^3$He vacua.

The same situation takes place for the relativistic quantum vacuum. This can be seen using  the nonlinear extension of the Hawking description of the quantum vacuum phenomenology in terms of the 4-form field.
In this description, the cosmological constant in Einstein equations is equivalent  to the thermodynamic potential,
 $\Lambda_{\rm vac}=\epsilon(q) -\mu q=-P_{\rm vac}$ \cite{KlinkhamerVolovik2008a,KlinkhamerVolovik2008b}, where $q$ is the  4-form field. 
 While $\epsilon(q)$ is determined by the microscopic trans-Planckian physics of the "atoms of the vacuum" and is very large,
 the pressure belongs to the infrared physics. That is why $\Lambda_{\rm vac}=0$ if our Universe  is isolated from the environment, as it happens for the $^3$He Universe  isolated from the environment. The Gibbs-Duhem thermodynamic identity ensures the cancellation of large vacuum energy in an equilibrium vacuum, regardless of the microscopic structure of the vacuum.

 \section{Standard Model of particle physics}

\subsection{Higgs bosons}

In the spin-triplet $p$-wave superfluid $^3$He with $3\times 3=9$ complex components of the order parameter the symmetry breaking gives rise to 18 collective bosonic modes. 
In $^3$He-B they are separated into  4  Nambu-Goldstone modes and 14 Higgs amplitude modes. 

\subsubsection{Nambu sum rule and hidden supersymmetry}

 Nambu \cite{Nambu1985} noticed that  in $^3$He-B these 18 modes, which are distributed into families with different angular momentum $J=0,1,2$, obey the rule valid for each family. In the relativistic form this Nambu sum rule reads:
  \begin{equation}
M_{J+}^2 + M_{J-}^2=4M_F^2
 \,.
\label{NambuRule}
\end{equation}
Here $M_+$ and $M_-$ are the gaps in the spectrum of modes, in which correspondingly the real and imaginary parts of the order parameter are oscillating; $M_F$ is the fermionic mass, which in $^3$He-B is the gap in fermionic spectrum of Bogoliubov quasiparticles.
This rule has been further extended to the thin film of the A-phase \cite{Zubkov2013,Zubkov2014}. The connection between the fermion and boson masses can be attributed to the manifestation of hidden supersymmetry in superfluid $^3$He \cite{Nambu1985}.

The Nambu sum rule is valid for the BCS weak coupling regime \cite{Sauls2017,NguenHalperin2019}. However, this rule can be applied to the extensions of the Standard Model, assuming that they are also in the weak coupling regime. In the simplest cases this  assumption suggests the existence 
of the second Higgs particle, with the mass 325 GeV in the $^3$He-B scenario or with the mass 245 GeV in the $^3$He-A scenario, in which $M_+=M_-$ \cite{Zubkov2013,Zubkov2014}.

\subsubsection{Little Higgs}

In $^3$He-B, due the tiny spin-orbit interaction one of the four Nambu-Goldstone modes becomes the Higgs boson with small mass, which is determined by spin-orbit coupling. This is the mode measured in experiments with the longitudinal NMR. The formation of the Higgs mass is the result of the violation of the $^3$He-B symmetry by spin-orbit interaction. In the modern language the enhanced symmetry which takes place when some terms in energy are small and are neglected, is called the hidden symmetry, and the corresponding Higgs with small mass is called pseudo Nambu-Goldstone boson.

The presently known Higgs boson has mass 125 GeV, which looks rather small compared to the typical electroweak energy scale of 1 TeV. That is why the natural guess from the $^3$He physics is that this boson is actually the pseudo Nambu-Goldstone mode \cite{Zubkov2015,Zavyalov2016,Khaidukov2017}, and one should search for the real Higgs, which could be as heavy as 1 TeV.

It is interesting that the hints of the Higgs bosons with 245 GeV, 325 GeV and with the TeV scale, have been reported
(see e.g. \cite{Meissner2013,ATLAS2012,Lane2016}), but not confirmed.

 \subsection{Nuclear Physics}
 
Connection between the chiral phenomena in the quark–gluon plasma and in the chiral superfluid $^3$He-A in relation to chiral magnetic and chiral vortical effects has been discussed in Sec. \ref{CMVE}. Here we consider another connection which is related to the models of hadrons and quark confinement.

 \subsubsection{MIT bag model and magnon condensate}
 
In $^3$He-B we can imitate the MIT bag model  for hadrons \cite{Autti2012a,Autti2012}. 
In this model \cite{Chodos1974a,Chodos1974b} the hadron is considered as  macroscopic box (bag) with the deconfinement phase inside, where the massless quarks are free, and with the vacuum in the confinement phase outside, where free quarks cannot exist. This model is similar to the electron bubble in a helium liquid, where the zero
point energy of electron in the ground state in the box potential compensates the external pressure and surface tension. In the MIT  bag the compensation also comes from the ground state energy of free quarks in the box. 

In $^3$He-B, we constructed the bosonic analog of the bag model: instead of the quarks in the ground state in a box potential, there are magnons which also fill the ground state forming the magnon Bose condensate \cite{Autti2012a,Autti2012}.

 \section{Quantum mechanics}
 
At the moment, $^3$He cannot say anything reasonable on the origin of quantum mechanics and quantum field theory. However, the possible origin of $\sqrt{-1}$ in quantum mechanics is suggested
\cite{Zubkov2014b,Zubkov2014c}. The microscopic physics of the quantum vacuum is fully described in terms of the real numbers, 
while the imaginary unit emerges in the low energy corner together with Weyl fermions. The reason for that is the same topology, which 
protects Weyl points in the chiral superfluid $^3$He-A.
 
 \section{Conclusion}
 
The Helium can answer most of the questions presented in the paper by Allen and Lidstr\"om "Life, the Universe, and
everything-42 fundamental questions" \cite{42questions}. At least Helium has opinion on these problems.
However, at the moment Helium cannot say anything physical on such questions as:
What is quantum mechanics? What is life? What is consciousness?
These topics still remain supernatural for Helium, and this is the reason why they
are not on the physical map in Fig. 1. The main task of Helium is to push forward the border line 
between the natural and supernatural parts of the Universe.

{\bf Acknowledgements}
This work has been supported by the European Research Council (ERC) under the European Union's Horizon 2020 research and innovation programme (Grant Agreement No. 694248).

\end{document}